\documentclass[reqno]{amsart}

\usepackage{amsmath}
\usepackage{amsfonts}
\usepackage{amsthm}

\numberwithin{equation}{section}
\newcommand {\pa}{\partial}

\newcommand{\Div}{{\operatorname{div}\,}}

\newcommand{\dist}{{\operatorname{dist}}}

\newcommand{\curl}{{\operatorname{curl}\,}}

\newtheorem{theorem}{Theorem}[section]

\newtheorem{lemma}[theorem]{Lemma}
\newtheorem{proposition}[theorem]{Proposition}

\def\Gg {{\mathcal G}} 
\def\R {{\mathbb R}} %R bold 
\def\N {{\mathbb N}} %N bold 
\def\Vg {{\mathcal V}} %V gothique
\title[Bulk superconductivity]{Bulk superconductivity in  Type II superconductors near the second
  critical field}

\author{S. Fournais}
\author{B. Helffer}

\address[S. Fournais]{Department of Mathematical Sciences, University of Aarhus, Ny Munke\-gade,
Building 1530,
DK-8000 Aarhus C, Denmark
}
\email{fournais@imf.au.dk}

\address[S. Fournais on leave from]{CNRS and Laboratoire de Math\'{e}matiques
  d'Orsay, Univ Paris-Sud, Orsay CEDEX, F-91405}

\address[B. Helffer]{Laboratoire de
Math\'{e}matiques UMR CNRS 8628\\ Universit\'{e} Paris-Sud - B\^{a}t 425\\
F-91405 Orsay Cedex\\ France.
}
\email{bernard.helffer@math.u-psud.fr}
\date{\today}

\begin{document}

\bibliographystyle{plain}

\begin{abstract}
We consider superconductors of Type II near the transition from the
'bulk superconducting' to the 'surface superconducting' state. We
prove a new $L^{\infty}$ estimate on the order parameter in the bulk, i.e. away from the boundary. This solves an open problem posed by
Aftalion and Serfaty \cite{AS}.
\end{abstract} 
\maketitle

%% \tableofcontents

\section{Introduction}
We consider a superconducting wire of cross section $\Omega \subset
\R^2$, which we assume to be regular and bounded.
The state of the material is described by 
the Ginzburg-Landau functional, which we write as
\begin{multline}
\label{eq:GL2D-def}
{\Gg}(\psi,{\bf A}) = \int_\Omega | p_{\kappa \sigma {\bf A}}\psi |^2
 -\kappa ^2| \psi |^2 +\frac {\kappa ^2}{2}| \psi |^4  \;
dx +   (\kappa \sigma)^2  \int_{\Omega}| \curl {\bf A}  - \beta | ^2\; dx\;.
\end{multline}
We use the notation $p_{{\bf A}} = (-i \nabla + {\bf A})$ for the
magnetic gradient.
In \eqref{eq:GL2D-def} $\kappa, \sigma$ are positive parameters, the wave function
(order parameter) $\psi$ describes the superconducting properties of
the given material and $(\kappa \sigma)\curl{\bf A}$ gives the induced magnetic
field. The function $(\kappa \sigma) \beta$ represents the external magnetic field
 and in this paper we will for simplicity consider the
case $\beta =1$, corresponding to  a constant external field of
intensity $\kappa \sigma$. We refer to \cite{deGeb,T} for a general introduction to the Physics of superconductivity and the Ginzburg-Landau model.

Consider the case $\sigma = \frac{\kappa}{b}$, with $b>0$. In the
limit $\kappa \rightarrow \infty$ (called Type II limit), the
following scenario presents itself. If $b < \Theta_0$, where $\Theta_0
\approx 0.59$ is a universal constant, the only minimizer of ${\Gg}$
(for large $\kappa$) is the state $(\psi=0, {\bf A} = {\bf F})$, where
$\curl {\bf F} =1$ \cite{LuPa1,He-Pan,FournaisHelffer2}. This is interpreted as the loss of
superconductivity for large external magnetic fields and the value of
$\sigma$ where this happens is denoted by $H_{C_3}$ and is called 'the
third critical field'.

Physicists consider a second critical 
field, $H_{C_2}$ which can be described
as follows---a precise definition being difficult to give. If $\sigma < H_{C_2}$, the material is in its
superconducting state in (a part of) the interior of the sample, whereas for $\sigma >
H_{C_2}$ superconductivity is restricted to a narrow region near the
boundary of $\Omega$. Different investigations show that---for large
values of $\kappa$---this
transition takes place near the value $\sigma = \kappa$, so even
though this critical field is difficult to define, one expects that
$H_{C_2}
\approx \kappa$ \cite{SS,Pan2}.

In this paper we will study what happens in the limit of large
$\kappa$---the Type II limit---when $\sigma =
\frac{\kappa}{b}$ with $b$ close to, but above the value $1$. In the
terminology of superconductivity, this means that we study the
parameter region close to but below the second critical field. In
this region the so-called Abrikosov lattices of vortices are supposed
to appear, but their description depends on a finer analysis than what
will be carried out in the present paper (see \cite{AS,Alm,SS} for
results in this direction). 

Our main result, Theorem~\ref{thm:ConjLinfty} below, gives for any
$\delta >0$ the
existence of a constant $C>0$ such that if $b>1$, $\kappa$ is large enough and $(\psi, {\bf A})$ is a
minimizer of ${\Gg}$ then
\begin{align}
  \label{eq:39}
  \| \psi \|_{L^{\infty}(\{ x \in \Omega \,: \, \dist(x,\partial
    \Omega)\geq \delta \} )} \leq C \sqrt{b-1}.
\end{align}
This implies tha,t in the interior of the sample, superconductivity is
weak in the uniform norm as $b$ approaches $1$.
Theorem~\ref{thm:ConjLinfty} thereby answers a question posed in \cite[p.~944
just below (1.20)]{SS} and more explicitly in \cite[List of open
problems p.7]{AS}.

Notice that  $|\psi|$ is not expected to become small at the boundary
when $b$ approaches the value $1$ \cite{Pan2,AlHe}.

Our interest in this problem was sparked by discussions with
S.~Serfaty. We would like to thank her for pointing our attention to
this interesting problem.

We end this introduction by discussing the optimality of the
estimate in  \eqref{eq:39}. 
According to \cite[Theorem~1.4]{SS}
(notice that the symbol $b$ in \cite{SS} denotes a different quantity
than in the present paper)
there exists a continuous, decreasing function $g:[0,1] \rightarrow
[0,1]$ such
that if $b\geq 1$, if $(\psi_{\kappa}, {\bf A}_{\kappa})_{\kappa \geq 1}$ denotes a
family of
minimizers of ${\mathcal G}$ with $\sigma = \frac{\kappa}{b}$ and if
$\{ B_{\kappa}\}_{\kappa \geq 1}$ is a family of balls such that 
\begin{enumerate}
\item $\kappa\,
\text{radius}(B_{\kappa}) \rightarrow \infty$,
\item $B_{\kappa} \cap \partial \Omega = 0$,
\end{enumerate}
then
\begin{align}
  \label{eq:40}
  \frac{1}{|B_{\kappa}|} \int_{B_{\kappa}} |\psi_{\kappa}|^4\,dx
  \rightarrow g(b^{-1}).
\end{align}
Furthermore, the function $g$ satisfies the double sided bound
\begin{align}
  \label{eq:41}
  \alpha (1-b^{-1})^2 \leq g(b^{-1}) \leq (1-b^{-1})^2,
\end{align}
for some $\alpha$: $0<\alpha <1$.

Combining \eqref{eq:40} and \eqref{eq:41} we see that $|\psi_{\kappa}|$
is of order $\sqrt{b-1}$ for $b$ near and above $1$ and in the
$L^4$-average sense given by \eqref{eq:40}. More precisely, (by taking
the balls $B_{\kappa}$ to be contained in $\{ x \in \Omega \,: \, \dist(x,\partial
    \Omega)\geq \delta \}$) we get the
lower bound 
\begin{align*}
\liminf_{\kappa \rightarrow \infty} \| \psi_{\kappa}
  \|^4_{L^{\infty}(\{ x \in \Omega \,: \, \dist(x,\partial
    \Omega)\geq \delta \})}&\geq
\liminf_{\kappa \rightarrow \infty} \| \psi_{\kappa}
  \|^4_{L^{\infty}(B_{\kappa})} \\
&\geq g(b^{-1}) \\
&\geq \alpha \frac{(b-1)^2}{b^2},
\end{align*}
complementary to \eqref{eq:39} thereby yielding the
optimality of the inequality.

\section{Uniform estimates on the Ginzburg-Landau system}
We will study solutions to the Ginzburg-Landau equations, i.e. the
stationary points of the GL-functional. For concreteness, let us
assume that $\Omega \subset \R^2$ is a bounded, smooth and simply
connected domain. These assumptions are likely to be unnecessarily
restrictive, but they cover the most interesting cases and allow us to work
without worrying about topological problems and regularity questions.

The Ginzburg-Landau equations are
\begin{subequations}
\label{eq:GL}
\begin{align}
\left.\begin{array}{c}
p_{\kappa \sigma {\bf A}}^2\psi =
\kappa^2(1-|\psi|^2)\psi \\
\label{eq:equationA}
\curl^2 {\bf A}  =-\tfrac{1}{\kappa \sigma} \Re\Big( \overline{\psi} \,p_{\kappa \sigma {\bf A}} \psi \Big)
\end{array}\right\} &\quad \text{ in } \quad \Omega \, ;\\
\left. \begin{array}{c}
\nu \cdot p_{\kappa \sigma{\bf A}} \psi = 0 \\
\curl {\bf A} = 1
\end{array} \right\} &\quad \text{ on } \quad \partial\Omega \, .
\end{align}
\end{subequations}
Here, for ${\bf A} = (A_1,A_2)$,  $\curl {\bf A} = \partial_{x_1}A_2 -
\partial_{x_2}A_1$, and
$$
\curl^2 {\bf A} =
(\partial_{x_2}(\curl {{\bf A}}),-\partial_{x_1}(\curl {\bf A})) \, .
$$
Using gauge invariance it is of no loss of generality to consider only
(weak) solutions $(\psi, {\bf A})$ of \eqref{eq:GL} satisfying that
$(\psi, {\bf A}) \in H^1(\Omega,\mathbb C)\times H^1_{\rm div}(\Omega)$, where
\begin{multline}
\label{eq:DefH1div}
H^1_{\rm div}(\Omega)= \Big\{ \Vg=(V_1,V_2)\in H^1(\Omega)^2\;\big|\; 
\Div {\mathcal V} = 0\,\mbox{ in } \Omega\,,\, \Vg \cdot \nu = 0\,\mbox{ on } \pa \Omega\Big\}\;.
\end{multline}
The space $H^1_{\rm div}(\Omega)$ inherits the topology (norm) from
$H^1(\Omega; \R^2)$.
We denote by ${\bf F}$ the unique vector potential in $H^1_{\rm
  div}(\Omega)$ with $\curl {\bf F} =1$.

When we want to stress for which values of the parameters $\kappa,
\sigma$ the system \eqref{eq:GL} is considered we will place these as
indices. For instance we will say that $(\psi,{\bf
  A})_{\kappa,\sigma}$ is a solution to \eqref{eq:GL}.

Recall (see for instance \cite{DGP} for a proof) that by the maximum principle we have the estimate 
\begin{align}
  \label{eq:31}
  \| \psi \|_{\infty} \leq 1,
\end{align}
for all solutions to \eqref{eq:GL}. 

We will in this paper use the notation $|t|_{+}$ for the 'positive
part', i.e. the function
\begin{align*}
  \R \ni t \mapsto |t|_{+} := \max(t,0).
\end{align*}

Our main result, the precise version of \eqref{eq:39}, is as follows.

\begin{theorem}\label{thm:ConjLinfty}~\\
There exists a constant $C^{(2)}_{\rm max}>0$ such that if $g_1:\R_+
\rightarrow \R_+$ with $g_1(\kappa) \rightarrow +\infty$, and 
$\frac{g_1(\kappa)}{\kappa} \rightarrow 0$ as $\kappa
\rightarrow \infty$ and
\begin{align}
  \label{eq:1}
  \omega_{\kappa}:=\{x \in \Omega \,:\, \dist(x,\partial \Omega) \geq
  \frac{g_1(\kappa)}{\kappa} \},
\end{align}
then there exists a function $g_2:\R_+
\rightarrow \R_+$ with $g_2(\kappa) \rightarrow 0$ as $\kappa
\rightarrow \infty$ such that
\begin{align}
  \label{eq:30}
  \| \psi \|_{L^{\infty}(\omega_{\kappa})} \leq C_{\rm max}^{(2)} \big|
  \frac{\kappa}{\sigma}-1\big|_{+}^{1/2} + g_2(\kappa),
\end{align}
for all solutions $(\psi, {\bf A})_{\kappa,\sigma}$ to \eqref{eq:GL} with $\kappa\geq 1$.
\end{theorem}

In the proof of Theorem~\ref{thm:ConjLinfty} we will use the {\it a priori} estimates
\begin{align}
  \label{eq:42}
  \| {\bf A} - {\bf F} \|_{W^{2,p}(\Omega)} \leq C_p \frac{1 + \kappa
    \sigma + \kappa^2}{\kappa \sigma} \| \psi \|_2 \|\psi \|_{\infty},
\end{align}
valid for all $\kappa, \sigma>0$ and all solutions of \eqref{eq:GL}
established in \cite[Equation (3.9)]{FournaisHelffer4}, and
\cite[Equation (3.15)]{FournaisHelffer4}:
\begin{align}
  \label{eq:46}
  \| \curl {\bf A} -1 \|_2 \leq \frac{C}{\sigma} \| \psi \|_{\infty}
  \| \psi \|_2.
\end{align}

\begin{proof}
By \eqref{eq:31} the statement for $\kappa >
2\sigma$ is obvious. On the other hand, by Giorgi-Phillips
\cite{Giorgi-Phillips} (see also \cite{FHBook}), if
\begin{align}
  \label{eq:32}
  \sigma\geq C_{GP} \max\{\kappa,1\},
\end{align}
then all solutions to \eqref{eq:GL} have $\psi=0$. Thus it suffices to consider the case
\begin{align}
  \label{eq:33}
  C^{-1} \kappa \leq \sigma \leq C \kappa, \quad\quad \kappa \geq 1.
\end{align}
Suppose for contradiction that \eqref{eq:30} is false. Then, for all
$N>0$ sufficiently large there exists a sequence $\{ (\psi_n, {\bf A}_n, \kappa_n,
\sigma_n) \}_{n \in \N}$ with $(\psi_n,{\bf A}_n)_{\kappa_n,
  \sigma_n}$ solution to \eqref{eq:GL} and such that $\kappa_n
\rightarrow \infty$ and
\begin{align}
  \label{eq:34}
  \| \psi_n \|_{L^{\infty}(\omega_{\kappa_n})} \geq N \big|
  \frac{\kappa_n}{\sigma_n}-1\big|_{+}^{1/2} + N^{-1}.
\end{align}
Due to \eqref{eq:33} we may assume, by possibly extracting a
subsequence, that
\begin{align}
  \label{eq:35}
  \frac{\kappa_n}{\sigma_n} \rightarrow b \in [C^{-1}, C].
\end{align}
Using \eqref{eq:42} and the compactness of the imbedding
$W^{2,p}(\Omega) \rightarrow C^{1,1/2}(\overline{\Omega})$ for $p>2$
we may assume---by possibly extracting a further subsequence---that
\begin{align}
  \label{eq:43}
  {\bf A}_n \rightarrow {\bf \tilde A}\qquad \text{ in } \quad C^{1,1/2}(\overline{\Omega}).
\end{align}
By \eqref{eq:46} we have
\begin{align}
  \label{eq:47}
  \curl {\bf \tilde A} = 1.
\end{align}
Let $P_n \in \omega_{\kappa_n}$ be a point with $|\psi_n (P_n)| = \|
\psi_n \|_{L^{\infty}(\omega_{\kappa_n})}$. By \eqref{eq:34} and
\eqref{eq:31} we therefore have
\begin{align}
  \label{eq:36}
  N^{-1} \leq |\psi_n (P_n)| \leq 1.
\end{align}
After extracting a subsequence we assume that
\begin{align}
  \label{eq:44}
  P_n \rightarrow P \in \overline{\Omega}.
\end{align}
We consider the scaled functions
\begin{align*}
{\bf a}_n(y) &:= \frac{{\bf A}_n(P_n +\frac{y}{\sqrt{\kappa_n \sigma_n}}) - {\bf A}_n(P_n)}{1/\sqrt{\kappa_n \sigma_n}},\\
\varphi_n(y) &:= e^{-i\sqrt{\kappa_n \sigma_n}{\bf A}_n(P_n)\cdot y} \psi_n(P_n +\frac{y}{\sqrt{\kappa_n \sigma_n}}).
\end{align*}
Let $R>0$.
Since $g_1(\kappa) \rightarrow +\infty$, ${\bf a}_n$, $\varphi_n$ are
defined on $B(0,R)$ for all $n$ sufficiently large.
The equation for $\psi$ in \eqref{eq:equationA} implies, since $\Div {\bf a}_n = 0$, that
\begin{align}
\label{eq:phi}
-\Delta \varphi_n - 2i {\bf a}_n \cdot \nabla \varphi_n + |{\bf a}_n|^2 \varphi_n
= \frac{\kappa_n}{\sigma_n} (1 -  |\varphi_n|^2)\varphi_n.
\end{align}
The convergence from \eqref{eq:43} and \eqref{eq:44} imply that 
\begin{align}
  \label{eq:45}
  {\bf a}_n(y) \rightarrow {\tilde F}(y) :=D{\bf \tilde A}(P) y,
\end{align}
with convergence in $C^{1/2}(B(0,R))$ for all $R>0$.
By \eqref{eq:47} we find
\begin{align}
  \label{eq:48}
  \curl {\bf \tilde F} =1.
\end{align}
The uniform (in $n$) boundedness of the coefficients to the equation \eqref{eq:phi}
for $\varphi_n$ implies boundedness of $\{ \varphi_n \} \subset
W^{2,p}(B(0,R/2))$ for all $p<\infty$ and all $R>1$. The compactness
of the imbedding $W^{2,p} \rightarrow C^{1}$ (for $p>2$) implies that we may
extract a convergent subsequence in $C^{1}(B(0,R/2))$. A diagonal sequence argument now
gives the existence of a limiting function
$\varphi \in L^{\infty}(\R^2)$ with 
\begin{align}
  \label{eq:38}
  N^{-1} + N |b-1|_{+}^{1/2} \leq \| \varphi \|_{L^{\infty}(\R^2)} \leq 1,
\end{align}
and
\begin{align}
  \label{eq:37}
  (-i\nabla +{\bf \tilde F})^2 \varphi = b (1-|\varphi|^2)\varphi.
\end{align}
Since $\curl {\bf \tilde F} =1$,
this is a contradiction to
Theorem~\ref{thm:Small} below if $N \geq C_{\rm max}$.
\end{proof}

\section{Estimates for the global problem}
We will consider the following equation of Ginzburg-Landau type,
\begin{align}\label{eq:GLb} \tag{$GL_b$}
 p_{\bf F}^2 u = b (1 - |u|^2) u, \quad\quad \text{ on } \R^2,
\end{align}
where $b \in \R$ is a parameter and ${\bf F}$ satisfies $\curl {\bf F}
=1$ in $\R^2$.
For concreteness we use the gauge freedom of the problem to fix the
choice
\begin{align*}
  {\bf F} = (-x_2/2,x_1/2).
\end{align*}
\begin{theorem}\label{thm:Small}~\\
(i)  If  $u \in
L^{\infty}(\R^2)$ is a solution to \eqref{eq:GLb} with  $b\leq 1$, then  $u=0$.\\
(ii) 
There exists a universal constant $C_{\rm max}>0$ such that if $u \in
L^{\infty}(\R^2)$ is a solution to \eqref{eq:GLb} with  $b>1$, then
\begin{align}
\label{eq:16}
  \| u \|_{\infty} \leq \min\{1, C_{\rm max} \sqrt{b-1}\}
\end{align}
\end{theorem}
It is well-known \cite{LuPa1,FournaisHelffer4,FHBook}
 that the equation
\eqref{eq:GLb} only admits trivial $L^{\infty}$-solutions\footnote{The case $b<1$ can be considered a magnetic special case of a Theorem by Sch'nol \cite{Sch'n,CFKS}.} if $b\leq 1$. 
Also,
it is a standard consequence of the maximum principle that bounded solutions
satisfy 
\begin{align}
  \label{eq:MaxPr}
  \| u \|_{\infty} \leq 1.
\end{align}
Thus only the second half of \eqref{eq:16} needs to be proved.

Define 
\begin{align}
  \label{eq:2}
  S(b) := \{ u \in L^{\infty}(\R^2) \,:\, u \text{ solves } \eqref{eq:GLb}\}.
\end{align}
and
\begin{align}
  \label{eq:3}
  M(b) := \sup_{ u \in S(b)} \| u \|_{\infty}.
\end{align}

The starting point is the following lemma.

\begin{lemma}\label{lem:First}
As $\epsilon \searrow 0$, we have the following estimate
\begin{align}
  \label{eq:6}
  M(1+\epsilon) = o(1).
\end{align}
\end{lemma}

\begin{proof}
The proof is by contraposition in the spirit of
\cite{FournaisHelffer4,LuPa1}. Suppose that Lemma~\ref{lem:First} is wrong. Then there exists
a sequence $\{ \epsilon_n \}_{n \in \N} \subset \R_+$ with $\epsilon_n
\rightarrow 0$ and an associated sequence $\phi_n$ of solutions to
$(GL_{1+\epsilon_n})$ with
\begin{align}
  \label{eq:7}
  \| \phi_n \|_{\infty} \geq \delta > 0.
\end{align}
Clearly, there will then exist a point $x_n \in \R^2$ with
$|\phi_n(x_n)| \geq \delta/2$. By magnetic translation invariance of
\eqref{eq:GLb} we may assume that $x_n = 0$ for all $n$.

By elliptic regularity and \eqref{eq:MaxPr}, $\{ \phi_n \}$ is bounded
in $W^{2,p}(B(N))$ for all $N \in \N$ and all $p<\infty$.
By compactness we may---for any given $s<2$, $p<\infty$ and $N \in
\N$---extract a convergent subsequence in $W^{s,p}(B(N))$.

By a diagonal sequence argument we get a $\phi \in W^{s,p}_{\rm
  loc}(\R^2)$ and a subsequence---still denotes by $\{ \phi_n \}$ such
that
\begin{align*}
  \| \phi_n - \phi \|_{W^{s,p}(B(N))} \rightarrow 0,
\end{align*}
for all $N$.
In particular, we see that $\| \phi \|_{\infty} \leq 1$, 
\begin{align}
  \label{eq:8}
  | \phi(0) | \geq \delta/2,
\end{align}
and $\phi$ solves $(GL_1)$. But we know from
\cite[Proposition~4.1]{FournaisHelffer4} (or part (i) of Theorem~\ref{thm:Small})
that the only bounded solution to $(GL_1)$ is $\phi=0$ in
contradiction to \eqref{eq:8}.
This finishes the proof of Lemma~\ref{lem:First}.
\end{proof}

\begin{proof}[Proof of Theorem~\ref{thm:Small}]~\\
Suppose for contradiction that a sequence of solutions $\{ \phi_n\}$
to $(GL_{1+\epsilon_n})$ exists with
\begin{align}
  \label{eq:10}
  \frac{\| \phi_n\|_{\infty}}{\sqrt{\epsilon_n}} \rightarrow \infty.
\end{align}

Define $\Lambda_n := \| \phi_n \|_{\infty}$. By magnetic translation
invariance, we may assume that $|\phi_n(0)| \geq
\frac{\Lambda_n}{2}$. Consider the function $f_n := \Lambda_n^{-1}
\phi_n$. This function satisfies $\| f_n \|_{\infty} \leq 1$ and
\begin{align}
  \label{eq:11}
  p_{\bf F}^2 f_n = b_n (1-\Lambda_n^2 |f_n|^2)f_n,
\end{align}
with $b_n:=1+\epsilon_n$. 
After possibly extracting a subsequence, we find 
\begin{align}
  \label{eq:12}
  f_n \rightarrow f \in W^{3/2,2}_{\rm loc}(\R^2) \hookrightarrow
  L^{\infty}_{\rm loc}(\R^2),
\end{align}
where $f$ satisfies
the lower bound
\begin{align}
  \label{eq:13}
  1/2 \leq |f(0)| \leq \| f \|_{\infty} \leq 1.
\end{align}
Using Lemma~\ref{lem:First} we get the limiting equation for $f$:
\begin{align}
  \label{eq:14}
  p_{\bf F}^2 f = f.
\end{align}
Thus $f$ lies in the lowest Landau band.

Let $\Pi_0$ be the projection on the lowest Landau band. This operator
is given explicitly by the integral kernel
\begin{align}
  \label{eq:5}
  \Pi_0(x,y) = \frac{1}{2\pi} e^{\frac{i}{2}(x_1y_2-x_2 y_1)} e^{-\frac{1}{2}(x-y)^2},
\end{align}
in particular, we see that $\Pi_0$ is a bounded operator on
$L^2(\R^2)$ and on $L^{\infty}(\R^2)$. By interpolation, $\Pi_0$ is
continuous on $L^p(\R^2)$ for all $p \in [2,\infty]$. 

The boundedness of $f_n$ and elliptic regularity applied to
\eqref{eq:11} imply that the conditions of
Proposition~\ref{prop:Formal} below are satisfied. Therefore, we get by
application of $\Pi_0$ to \eqref{eq:11} that
\begin{align}
  \label{eq:9}
  0 = \Pi_0\{ \frac{\epsilon_n}{\Lambda_n^2} -|f_n|^2) f_n \}(x), \qquad \text{ for all }
  \quad x \in \R^2.
\end{align}
Using \eqref{eq:10} and passing to the limit in \eqref{eq:9} using \eqref{eq:5} and
dominated convergence, we obtain
\begin{align}
  \label{eq:15}
  \Pi_0\{ |f|^2 f \} = 0.
\end{align}
By Proposition~\ref{prop:non-exist} below we therefore conclude that
$f=0$, which is in contradiction to \eqref{eq:13}.
\end{proof}

\begin{proposition}\label{prop:Formal}
Suppose that $f, p_{\bf F}^2 f \in L^{\infty}(\R^2) \cap
C(\R^2)$. Then
\begin{align*}
  \big(\Pi_0 (p_{\bf F}^2 -1) f\big)(x) = 0, \qquad \text{ for all }
  \quad x \in \R^2.
\end{align*}
\end{proposition}

\begin{proof}
By continuity of $f$, boundedness of $f$ and gaussian decay of the
kernel of $\Pi_0$ we get that $\Pi_0f$ is continuous. The same
argument applies to $\Pi_0 (p_{\bf F}^2 f)$ and therefore $\Pi_0 (p_{\bf
  F}^2 -1 ) f$ is continuous.
Therefore, it suffices to prove that
\begin{align*}
  \int \varphi(x) \big(\Pi_0 (p_{\bf F}^2 -1) f\big)(x) \,dx = 0,
\end{align*}
for all $\varphi \in C^{\infty}_0(\R^2)$, which is immediate.
\end{proof}

\begin{proposition}\label{prop:non-exist}~\\
Suppose that $f \in L^{\infty}(\R^2)$ satisfies
\begin{align}
  \label{eq:27Old}
  p_{\bf F}^2 f = f, \quad \text{ and }\quad \Pi_0(|f|^2 f) = 0,
\end{align}
then $f=0$.
\end{proposition}

Below we will use the localization functions $\chi_R$ defined as
follows. Let $\chi \in C^{\infty}(\R)$ be even, non-increasing on $\R_{+}$ and satisfy
\begin{align}
  \label{eq:28}
  \chi(t)=1\quad \text{ for } |t|\leq 1,\quad\quad\quad
  \chi(t)=0\quad \text{ for } |t|\geq 3/2.
\end{align}
Define, for $R>0$, $x\in \R^2$,
\begin{align}
  \label{eq:29}
  \chi_R(x) := \chi(|x|/R).
\end{align}
\begin{proof}[Proof of Proposition~\ref{prop:non-exist}]~\\
Since $f \in L^{\infty}({\mathbb R}^2)$, we clearly have
\begin{align}
  \label{eq:25}
  \int_{\{|x|\leq R\}} |f(x)|^4\,dx \leq C R^2,
\end{align}
for all $R>0$. We will prove that one can recursively improve the
power of $R$ in \eqref{eq:25}, i.e. if the estimate
\begin{align}
  \label{eq:26}
  \int_{\{|x|\leq R\}} |f(x)|^4\,dx \leq C R^s,
\end{align}
holds for all $R>1$ and some constant $C$, then there exists a new
constant $C'$ such that
\begin{align}
  \label{eq:27}
  \int_{\{|x|\leq R\}} |f(x)|^4\,dx \leq C' R^{s-\frac{1}{2}},
\end{align}
for all $R>1$.

Since we get a negative power of $R$ after a finite number of steps
that will imply that $f=0$. Thus we only need to prove that
\eqref{eq:27} follows from \eqref{eq:26}.

We calculate, using $(p_{\bf F}^2 - 1) f = 0$,
\begin{align}
  \label{eq:19}
 \langle (p_{\bf F} ^2 -1)\chi_R f\,|\, \chi_R f\rangle  &=
 \| (\nabla \chi_R) \; f\|_2^2 \leq \frac{C}{R^2} \int_{\{|x|\leq
   2R\}} |f|^2\,dx \nonumber \\
&\leq \frac{C'}{R} \sqrt{\int_{\{|x|\leq
   2R\}} |f|^4\,dx} \leq C'' R^{\frac{s}{2}-1}.
\end{align}
This gives, by $L^2$-projection, and dropping the primes on the constant
\begin{align}
  \label{eq:17}
  \|\Pi_0^{\perp} (\chi_R f) \|_2^2 \leq C R^{\frac{s}{2}-1} \;,
\end{align}
where we have introduced the notation
$\Pi_0^{\perp}:=1-\Pi_0$.
Using that $\Pi_0$ is bounded from $L^{\infty}$ to $L^{\infty}$ we get
that $\|\Pi_0^{\perp} (\chi_R f) \|_{\infty} \leq C$, and by
interpolation
\begin{align}
  \label{eq:20}
  \|\Pi_0^{\perp} (\chi_R f) \|_4 \leq \|\Pi_0^{\perp} (\chi_R f)
  \|_2^{\frac{1}{2}}\|\Pi_0^{\perp} (\chi_R f)
  \|_{\infty}^{\frac{1}{2}} \leq C'R^{\frac{s}{4}-\frac{1}{2}}.
\end{align}

We now write
\begin{align*}
  \int \chi_R |f|^4\,dx &= \int \overline{\chi_R f} \, \Pi_0^{\perp}
  (|f|^2 f)\,dx \\
&= \int \overline{\chi_R f} \,\Pi_0^{\perp} \chi_{2R} (|f|^2 f)\,dx +
\int \overline{\chi_R f} \,\Pi_0^{\perp} (1-\chi_{2R}) (|f|^2 f)\,dx \\
&= \langle \Pi_0^{\perp}(\chi_R f)\, | \,\chi_{2R} (|f|^2 f) \rangle -
\int \overline{\chi_R f}\, \Pi_0 (1-\chi_{2R}) (|f|^2 f)\,dx.
\end{align*}
Here we used that $\Pi_0^{\perp} = 1 - \Pi_0$ and that $\chi_R
(1-\chi_{2R})=0$ to get the last identity.

By H\"{o}lder's inequality combined with \eqref{eq:20} and
Lemma~\ref{lem:localityPi0} below, we can therefore
estimate
\begin{align}
  \label{eq:18}
    \int \chi_R |f|^4\,dx &\leq  \|\Pi_0^{\perp} (\chi_R f) \|_4 \|
    \chi_{2R} |f|^2 f \|_{4/3} + Ce^{-\frac{1}{16}R^2} \nonumber \\
&\leq C R^{\frac{s}{4}-\frac{1}{2}} R^{\frac{3}{4}s}  + Ce^{-\frac{1}{16}R^2}.
  \end{align}
Therefore, we get for some new constant $C>0$,
\begin{align}
  \label{eq:21}
  \int_{\{|x|\leq R\}} |f|^4\,dx \leq C R^{2-\frac{1}{2}},
\end{align}
which is \eqref{eq:27}. This finishes the proof.
\end{proof}

\begin{lemma}\label{lem:localityPi0}~\\
  There exists a constant $C>0$ such that
  \begin{align*}
    \int_{{\mathbb R}^2} \{(1-\chi_{2R}) \Pi_0 \chi_R u\}\, v\,dx \leq C e^{-R^2/16},
  \end{align*}
  for all $u,v\in L^{\infty}({\mathbb R}^2)$ with $\|u\|_{\infty}, \|v \|_{\infty} \leq 1$ and all $R>1$.
\end{lemma}

\begin{proof}[Proof of Lemma~\ref{lem:localityPi0}]~\\
Upon inserting the explicit integral kernel of $\Pi_0$, we get
\begin{align*}
  \Big| \int \{(1-\chi_{2R}) \Pi_0 \chi_R u\}\, v\,dx \Big|
&\leq
C \int_{\{|x|\leq 3R/2\}} \int_{\{|y|\geq 2R\}} e^{-|x-y|^2/2}\,dxdy \\
&\leq
C' R^2 \int_{\{|y|\geq 2R\}} e^{-(|y|-3R/2)^2/2}\,dy,
\end{align*}
from which the estimate is immediate.
\end{proof}

\medskip\par
\noindent{\bf Acknowledgements}\\
The two authors were partially supported by the ESF
Scientific Programme in Spectral Theory and Partial Differential
Equations (SPECT).
SF is supported by a Skou Grant and a Young Elite Researcher Award
from the Danish Research Council.

\end{document}